\documentclass[twocolumn, prl,aps,showpacs,superscriptaddress]{revtex4}
\usepackage{graphics}
\usepackage{url}
\begin{document}

\newcommand\Hbar{$\overline{\rm H}$}
\newcommand\Hbars{$\overline{\rm H}$s}
\newcommand\pbar{$\overline{\rm p}$}
\newcommand\pos{e$^+$}
\newcommand\elec{e$^-$}
\newcommand\us{$\,$}

\title{Compression of Antiproton Clouds for Antihydrogen Trapping}
\author{G.B. Andresen}
\affiliation{Department of Physics and Astronomy, Aarhus University, DK-8000 Aarhus C, Denmark}
\author{W. Bertsche}
\affiliation{Department of Physics, Swansea University, Swansea SA2 8PP, United Kingdom}
\author{P.D. Bowe}
\affiliation{Department of Physics and Astronomy, Aarhus University, DK-8000 Aarhus C, Denmark}
\author{C.C. Bray}
\affiliation{Department of Physics, University of California at Berkeley, Berkeley, CA 94720-7300, USA}
\author{E. Butler}
\affiliation{Department of Physics, Swansea University, Swansea SA2 8PP, United Kingdom}
\author{C.L. Cesar}
\affiliation{Instituto de F\'{i}sica, Universidade Federal do Rio de Janeiro, Rio de Janeiro 21941-972, Brazil}
\author{S. Chapman}
\affiliation{Department of Physics, University of California at Berkeley, Berkeley, CA 94720-7300, USA}
\author{M. Charlton}
\affiliation{Department of Physics, Swansea University, Swansea SA2 8PP, United Kingdom}
\author{J. Fajans}
\affiliation{Department of Physics, University of California at Berkeley, Berkeley, CA 94720-7300, USA}
\author{M.C. Fujiwara}
\affiliation{TRIUMF, 4004 Wesbrook Mall, Vancouver BC, Canada V6T 2A3}
\author{R. Funakoshi}
\affiliation{Department of Physics, University of Tokyo, Tokyo 113-0033, Japan}
\author{D.R. Gill}
\affiliation{TRIUMF, 4004 Wesbrook Mall, Vancouver BC, Canada V6T 2A3}
\author{J.S. Hangst}
\affiliation{Department of Physics and Astronomy, Aarhus University, DK-8000 Aarhus C, Denmark}
\author{W.N. Hardy}
\affiliation{Department of Physics and Astronomy, University of British Columbia, Vancouver BC, Canada V6T 1Z4}
\author{R.S. Hayano}
\affiliation{Department of Physics, University of Tokyo, Tokyo 113-0033, Japan}
\author{M.E. Hayden}
\affiliation{Department of Physics, Simon Fraser University, Burnaby BC, Canada V5A 1S6}
\author{R. Hydomako}
\affiliation{Department of Physics and Astronomy, University of Calgary, Calgary AB, Canada T2N 1N4}
\author{M.J. Jenkins}
\affiliation{Department of Physics, Swansea University, Swansea SA2 8PP, United Kingdom}
\author{L.V. J\o rgensen}
\affiliation{Department of Physics, Swansea University, Swansea SA2 8PP, United Kingdom}
\author{L. Kurchaninov}
\affiliation{TRIUMF, 4004 Wesbrook Mall, Vancouver BC, Canada V6T 2A3}
\author{R. Lambo}
\affiliation{Instituto de F\'{i}sica, Universidade Federal do Rio de Janeiro, Rio de Janeiro 21941-972, Brazil}
\author{N. Madsen}
\affiliation{Department of Physics, Swansea University, Swansea SA2 8PP, United Kingdom}
\author{P. Nolan}
\affiliation{Department of Physics, University of Liverpool, Liverpool L69 7ZE, United Kingdom}
\author{K. Olchanski}
\affiliation{TRIUMF, 4004 Wesbrook Mall, Vancouver BC, Canada V6T 2A3}
\author{A. Olin}
\affiliation{TRIUMF, 4004 Wesbrook Mall, Vancouver BC, Canada V6T 2A3}
\author{A. Povilus}
\affiliation{Department of Physics, University of California at Berkeley, Berkeley, CA 94720-7300, USA}
\author{P. Pusa}
\affiliation{Department of Physics, University of Liverpool, Liverpool L69 7ZE, United Kingdom}
\author{F. Robicheaux}
\affiliation{Department of Physics, Auburn University, Auburn, AL 36849-5311, USA}
\author{E. Sarid}
\affiliation{Department of Physics, NRCN-Nuclear Research Center Negev, Beer Sheva, IL-84190, Israel}
\author{S. Seif El Nasr}
\affiliation{Department of Physics and Astronomy, University of British Columbia, Vancouver BC, Canada V6T 1Z4}
\author{D.M. Silveira}
\affiliation{Instituto de F\'{i}sica, Universidade Federal do Rio de Janeiro, Rio de Janeiro 21941-972, Brazil}
\author{J.W. Storey}
\affiliation{TRIUMF, 4004 Wesbrook Mall, Vancouver BC, Canada V6T 2A3}
\author{R.I. Thompson}
\affiliation{Department of Physics and Astronomy, University of Calgary, Calgary AB, Canada T2N 1N4}
\author{D.P. van der Werf}
\affiliation{Department of Physics, Swansea University, Swansea SA2 8PP, United Kingdom}
\author{J.S. Wurtele}
\affiliation{Department of Physics, University of California at Berkeley, Berkeley, CA 94720-7300, USA}
\author{Y. Yamazaki}
\affiliation{Atomic Physics Laboratory, RIKEN, Saitama 351-0198, Japan}
\collaboration{ALPHA Collaboration}
\noaffiliation

\date{Received \today}

\begin{abstract} Control of the radial profile of trapped antiproton clouds is critical to trapping antihydrogen.
We report the first detailed measurements of the radial manipulation of antiproton clouds, including areal density compressions by factors as large as ten, by manipulating spatially overlapped electron plasmas.   We show detailed measurements of the near-axis antiproton radial profile and its relation to that of the electron plasma.
\end{abstract}


\maketitle

\begin{figure*}[t]
\centerline{\resizebox{16cm}{!}{\includegraphics{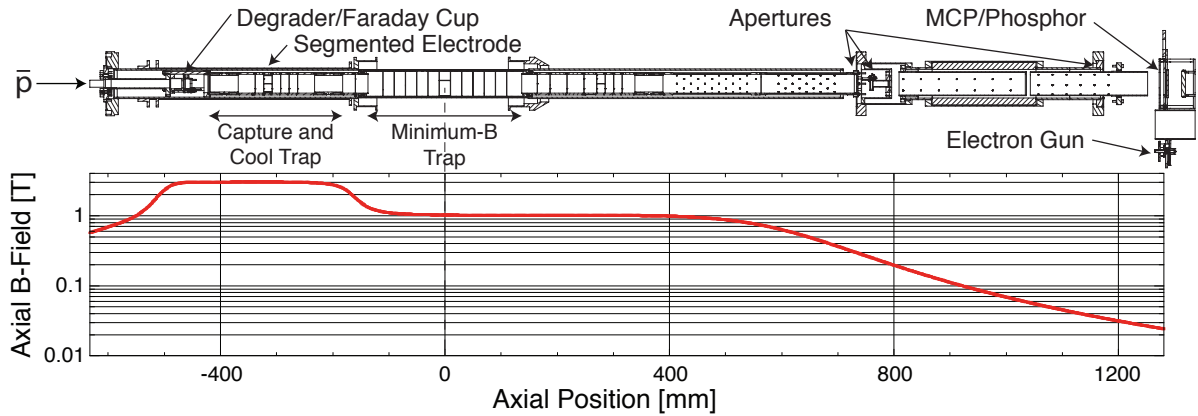}}}
\caption{Schematic diagram of the ALPHA apparatus.    A moveable probe on the right alternately inserts an electron gun and a MCP/Phosphor assembly.  The graph below the schematic plots the axial magnetic field in the trap.}
\label{apparatus}
\end{figure*}

Cold antihydrogen atoms (\Hbar) were first produced by the ATHENA collaboration \cite{amor:02}, and, shortly thereafter, by ATRAP \cite{gabr:02} at the CERN Antiproton Decelerator (AD) \cite{maur:97} in 2002.  They were produced by mixing positrons (\pos) and antiprotons (\pbar) held in Penning-Malmberg traps.  Such traps use an axial magnetic field to provide radial confinement and electrostatic wells to provide axial confinement.  Penning-Malmberg traps confine only charged particles and, consequently, do not confine neutral \Hbar\ atoms.  All the \Hbar\ atoms produced to-date have ionized in the electrostatic well fields or annihilated on the trap walls immediately after their formation.

The current generation of experiments \cite{andr:07,gabr:07} aims to trap  \Hbar\ atoms as this is likely necessary for precision CPT and gravity tests.  Neutral \Hbar\ atoms have a small permanent magnetic moment and can be trapped by a local, three-dimensional minimum of a magnetic field \cite{prit:83}.  Traps based on this effect are called Minimum-B traps.  To trap both charged and neutral species simultaneously, the Minimum-B and Penning-Malmberg traps must be co-located.  The compatibility of Minimum-B and Penning-Malmberg traps remains controversial \cite{faja:05,andr:07,gabr:07}, but it is clear that the two are most compatible if the \pbar's and \pos's are held close to the trap axis where the perturbations from the Minimum-B trapping fields are smallest \cite{faja:04,faja:06,zhan:07}. Furthermore, holding the \pbar's and \pos's near the axis increases their overlap, and slows the $\mathbf{E}\times\mathbf{B}$ drifts of the \pbar's.  These drifts increase the kinetic energy of the \Hbar's that form from the \pbar's, and make the \Hbar's more difficult to confine in the very shallow Minimum-B traps.

Successful \pbar\ compression has been briefly reported elsewhere \cite{yama:01a,kuro:05,tori:05};  here we present the first carefully controlled and quantitative characterization of the process, as well as the first accurate measurements of the near-axis radial distribution of \pbar's.  We have demonstrated areal density increases by as much as a factor of ten, and have produced \pbar\ clouds with radii as small as 0.29\us mm.  Clouds of this size are far from the loss limits \cite{faja:06} of our trap \cite{andr:07}, and promise to be much easier to confine.

A schematic drawing of our apparatus is shown in Fig~\ref{apparatus}.  The experimental cycle begins by injecting $\sim 120$ million electrons (\elec's) into a 136\us mm long electrostatic well located in the 3\us T Capture and Cool trap.  The \elec's are later used to cool the \pbar's \cite{gabr:89}, and form a plasma with a radius of $\sim 0.84$\us mm.  They quickly cool (calculated energy e-folding time of 0.44\us s) via cyclotron radiation in the 3\us T field to near the temperature of the cryogenically cooled trap walls of diameter 33.6\us mm \cite{hyat:87,beck:96}.  We then adjust the radius of this plasma by applying a rotating electrostatic potential to an azimuthally segmented electrode (see Fig.~\ref{apparatus}). This technique, called a rotating wall (RW) potential \cite{huan:97,grea:00}, is commonly used in non-neutral plasma research to adjust the plasma density and radius, and works by applying a torque to a plasma.  If the RW frequency is higher than the plasma rotation frequency, this torque will compress the plasma; if it is lower, it will expand the plasma. (The plasma rotates because the radial electric field in the trap engenders an azimuthal $\mathbf{E}\times\mathbf{B}$ drift.)

After the \elec\ are injected and their radial profile is adjusted to optimize \pbar\ cooling, as discussed later, we compress them axially into a well of length 30\us mm by manipulating the electrostatic potentials. Next, we inject a pulse of $\sim 3\times 10^7$ antiprotons from the AD. The \pbar's pass through $\sim 218\,\mu$m aluminium-equivalent degrading foils so that $\sim 3\times 10^4$ have an energy of less than 5\us keV \cite{amor:04a}. These relatively slow \pbar's are captured by directing them into a one-sided, 5\us kV, electrostatic well.  Before they bounce out of this well, we erect a second 5\us kV wall at the entrance end of the trap, thereby confining them in the now complete well \cite{gabr:86,amor:04a}.  These hot \pbar's then cool by collisions with the \elec\ \cite{gabr:89,amor:04a} for 30\us s, after which the well potential walls are lowered to our working voltages of 10-100\us V.

\begin{figure}
\centerline{\resizebox{3.4in }{!}{\includegraphics{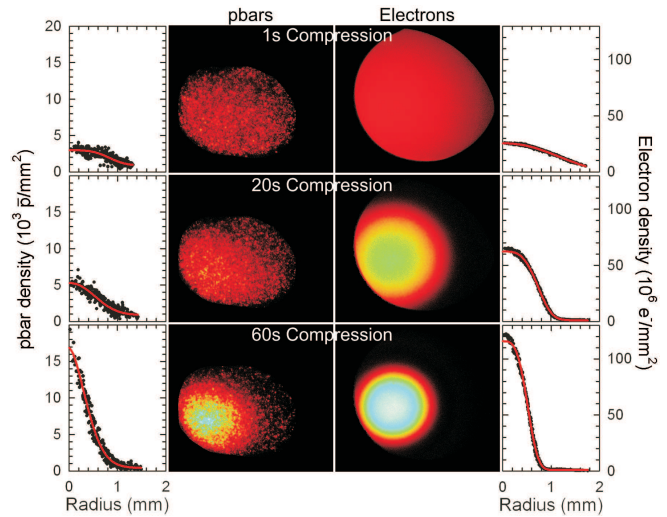}}}
\caption{\pbar\ and \elec\ images showing the effects of compression, and the resulting radial profiles.  The red lines are Gaussian-like (i.e.\ $\exp(-|r/r_0|^k)$, where $k\approx 2$) fits to the radial profiles. }
\label{compression_images}
\end{figure}

We image the \elec's and \pbar's by allowing them to escape along the magnetic field lines onto a microchannel plate (MCP)/phosphor screen assembly.  We capture the resultant images with a CCD camera \cite{peur:93a}.  Some typical images, described in detail below, are shown in Fig.~\ref{compression_images}.  The spatial distribution of the particles in the trap can be deduced by mapping the images back from the MCP to the particle trap.   Since the MCP is in a B-field of 0.024T, which is much lower than the trap field of 3T, the field lines, and, hence, the particles, expand by a factor of $\sqrt{3/0.024}=11.2$.  The \elec's are tightly bound to the field lines and follow them closely.  However, the heavier \pbar's exhibit small drifts, most notably the centrifugal drift \cite{chen:84} in the low magnetic field region near the MCP.  These drifts cause the \pbar's to rotate about the magnetic axis during extraction.  There are several apertures in our apparatus located near the positions indicated in Fig.~\ref{apparatus}; these apertures are clearly visible in Fig.~\ref{compression_images} and limit the maximum size of the plasmas that we can image. Because the \pbar's drift, and because the trap's magnet and mechanical axes are not perfectly aligned, the apertures are imaged differently for the two species.  As a result, the aperture image centers are not coincident, and the apertures limit the image area differently; the \pbar\ aperture image area is about 40\% smaller than the \elec\ aperture image area.

We calibrate the image brightness by independently measuring the charge with a Faraday cup (\elec's) and with scintillators (\pbar's).  The brightness is linearly \cite{peur:93a} related to the charge, and the calibrations are accurate to about 20\%. As there are far fewer \pbar's than \elec's, we operate the MCP at higher gain ($\sim 3\times 10^4$) for \pbar's than for \elec's ($\sim 300$).  Before imaging the \pbar's we extract all the \elec's by momentarily lowering the trap wall, thereby allowing the light \elec's to escape before the heavy \pbar's have time to react \cite{amor:04a}.  We repeat these ``e-kick'' cycles many times to ensure that we remove all the \elec's, while simultaneously monitoring annihilations to verify that we do not also lose \pbar's.  When we do not load \pbar's from our trap (by blocking the incoming \pbar\ beam, or by not applying the 5\us kV catching potentials), but otherwise run a normal cycle, we observe a null image.

\begin{figure}[t]
\centerline{\resizebox{2.5in}{!}{\includegraphics{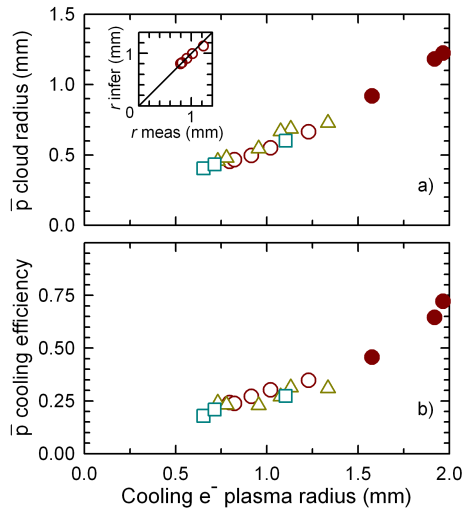}}}
\caption{a) The cooled \pbar\ cloud radius as a function of the radius of the \elec\ plasma used to cool the \pbar's.  The various symbols correspond to trials with differing total electron numbers (100-165\us M).  The \elec\ plasma radii for the hollow symbols were measured directly from the images; the radii for the filled symbols were calculated from the central intensity as described in the text. The inset figure shows the measured and inferred values for the points in the circle dataset where both methods could be employed.  b) The cooling efficiency (see text) as a function of the \elec\ plasma radius.}
\label{cooling}
\end{figure}

Figure \ref{cooling} shows the effect of varying the radius of the \elec\ plasma on the \pbar\ cooling process. In Fig.~\ref{cooling}a, we establish that the radius of the cooled \pbar\ cloud scales with the radius of the cooling \elec\ plasma.  We vary the \elec\ plasma radius by varying the RW frequency.  Before the RW is applied, the \elec\ radius is about 0.80\us mm, and is unchanged by a RW frequency of 400\us kHz.  We can compress the plasma to radii as small as 0.65\us mm and expand it to 1.95\us mm by applying RW frequencies that are, respectively, above (3\us MHz) and below (10\us kHz) this frequency.  Above a radius of about 1.39\us mm we do not observe enough of the \elec\ plasma edge to accurately measure its half-width-half-maximum [HWHM (our definition of the radius.)]  For these plasmas we assume that the \elec\ plasma radial profiles are self-similar, and infer the plasma radius from the peak density.  The inset graph in Fig.~\ref{cooling}a validates this approach.

In Fig.~\ref{cooling}b we plot the \pbar\ cooling efficiency as a function of the \elec\ plasma radius. (Less detailed measurements of this quantity have been reported by ASACUSA\cite{kuro:05}.) The cooling efficiency is the ratio of the number of cooled \pbar's to the total number of captured \pbar's, as measured by separately dumping these two populations into the degrader and recording the number of annihilations.  If we do not expand the \elec\ plasma with the RW, we cool only about 24\% of the captured \pbar's.  With expansion to about 1.95mm, we can cool up to about 72\%.  Computer simulations using the MAD \cite{grot:89} and SRIM \cite{zieg:85web} packages predict that the radius (sigma) of the incoming  \pbar's is about 4\us mm, which is compatible with our measurements when one considers that the plasmas extend beyond the HWHM, and the uncertainties in the simulations and their input parameters.

\begin{figure}
\centerline{\resizebox{2.5in}{!}{\includegraphics{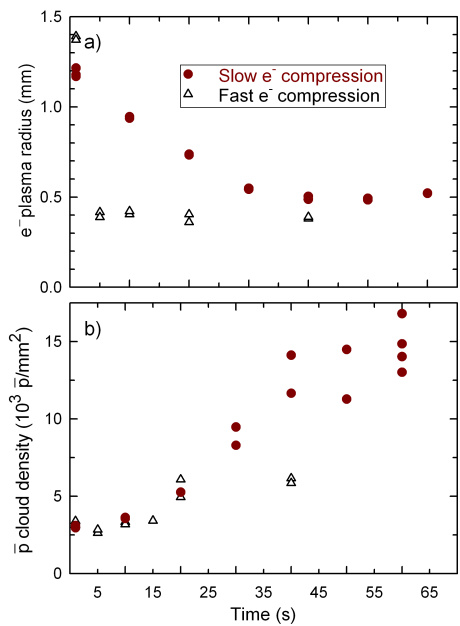}}}
\caption{The a) \elec\ radius and b) \pbar\ density as a function of time, for fast and slow compression.  Note that the \pbar\ density does not track the \elec\ compression if the latter is fast.}
\label{time_dependence}
\end{figure}

\begin{figure}
\centerline{\resizebox{2.5in}{!}{\includegraphics{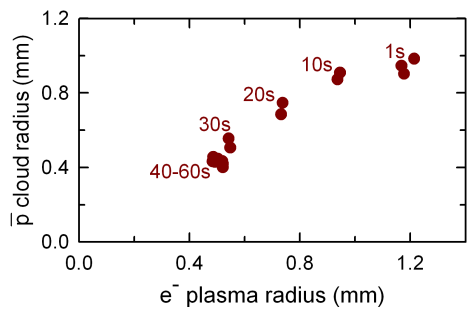}}}
\caption{The \pbar\ cloud radius as a function of the \elec\ plasma radius for the slow compression data shown in Fig.~\ref{time_dependence}.}
\label{radius_vs_radius}
\end{figure}

Figure~\ref{cooling} shows that efficient \pbar\ cooling requires \elec\ plasmas which are several millimeters in radius, and that the resulting \pbar\ clouds will be of comparable radius.  This radius is larger than optimal for the Minimum-B trap, particularly as the \pbar's expand by a factor of $\sim 1.7$ when they are transferred from the 3\us T Capture and Cool trap to the 1\us T Minimum-B trap (see Fig.~\ref{apparatus}).  Consequently, we use a second RW cycle, now operating on the mixed \elec -\pbar\  plasma, to compress the \elec's and \pbar's before effecting the transfer.  Typical results of this compression cycle are shown in Fig.~\ref{compression_images}; systematic \pbar\ compression studies are shown in Figs.~\ref{time_dependence} and \ref{radius_vs_radius}.  For the expansion RW cycle we used a chirped frequency drive, and controlled the radius and density by changing the final frequency.  For compression we could not use a chirped drive because of unwanted resonances in our system; instead, we used a single high frequency drive (10\us MHz for the data in Figs.~\ref{compression_images}, \ref{time_dependence} and \ref{radius_vs_radius}), and controlled the \elec radius by varying the time that the RW was applied.  As can be seen from Figs.~\ref{compression_images} and  \ref{radius_vs_radius}, the \pbar\ density and radius follow those of the \elec's;  the results are shown in Figs.~\ref{time_dependence}b and \ref{radius_vs_radius}.  In this series, the \pbar\ density increased by a factor of 5, and the radius decreased to about 0.42\us mm; about 11000 \pbar's were compressed. Using a higher frequency RW drive (20-25\us MHz) yielded \pbar\ densities twice as high and radii of 0.29\us mm.

Our measurements suggest that the \pbar's come into equilibrium with the \elec's, and that this equilibrium drives the \pbar\ cloud radius towards the \elec\ plasma radius.  Presumably, the charges interact through collisionally-mediated drag forces.  Such ``sympathetic'' compression has been observed in laser controlled multi-species ion plasmas \cite{lars:86}.  We commonly observe that when we compress the \elec's too quickly, the \pbar's do not follow the \elec's. (See, for instance, the fast \elec-compression data in Fig.~\ref{time_dependence}, where the compression speed was increased by increasing the RW drive voltage by a factor of 5.)  With fast \elec-compression, it is likely that the \pbar's are left behind in a region of low \elec\ density where the interspecies collision rate is too low to keep the species coupled.   Though the \pbar's do not compress if we eject the \elec's before the compression RW cycle, we cannot rule out the possibility that the RW compression acts directly on the \pbar's when the RW is applied to a mixed \elec-\pbar\ plasma.

For the density and temperature conditions of our \elec's and \pbar's, the global thermal equilibrium condition should place the \pbar's in a ring just outside the \elec plasma \cite{onei:81}.  Such distributions have been observed in laser cooled plasmas \cite{lars:86}, but we do not observe them.  We do not know if this is because the \pbar's have not yet fully relaxed, they are redistributed during the e-kicking process, the \elec\ plasmas are substantially hotter than we believe them to be, the imaging system smears our \pbar\ images, or something more fundamental is responsible.

In conclusion, we report the first detailed measurements of trapped \pbar\ radial compression. We can compress the \pbar\ density by a factor of ten and decrease the radii to 0.29\us mm. These clouds are 10-20 times smaller in radius than the clouds reported by ATHENA \cite{fuji:04} and ATRAP \cite{oxle:04}.  Control of the radial profile of the \pbar's is critical to their survival in a Minimum-B trap.  We have also studied the effect of the \elec\ plasma radius on the cooling of hot \pbar's.  Finally, we have developed a diagnostic that gives a detailed radial profile of \pbar's near the trap axis.  In the crucial near-axis region, this new diagnostic is a marked improvement over methods based upon annihilation imaging \cite{fuji:04}, or two data point extrapolations \cite{oxle:04}.

This work was supported by CNPq, FINEP (Brazil), ISF (Israel), MEXT (Japan), FNU (Denmark), NSERC, NRC/TRIUMF (Canada), DOE, NSF (USA), EPSRC and the Leverhulme Trust (UK) and HELEN/ALFA-EC.


\end{document}